\documentclass[aps,pra,groupedaddress,twocolumn]{revtex4-2}

\usepackage{graphicx}
\usepackage{dcolumn}
\usepackage{amsmath}
\usepackage{bm}

\begin{document}



\title{Light Shift Suppression in Coherent-Population-Trapping Atomic Clocks\\in the Field of Two Circularly Polarized Light Beams}


\author{D.V. Brazhnikov$^{1,2}$}
\email[Corresponding author: ]{x-kvant@mail.ru}
\author{S.M. Ignatovich$^1$}
\author{M.N. Skvortsov$^1$}
\affiliation{$^1$Institute of Laser Physics SB RAS, 15B Lavrentyev Avenue, Novosibirsk 630090, Russia}
\affiliation{$^2$Novosibirsk State University, 1 Pirogov Street, Novosibirsk 630090, Russia}

\date{\today}

\begin{abstract}
The state-of-the-art miniature atomic clocks (MACs) are based on the phenomenon of coherent population trapping (CPT) in alkali-metal atomic vapors (Rb or Cs). Increasing frequency stability of the clocks is an urgent issue that will lead to significant progress in many fields of application. Here, we examine a light field configuration composed of two bichromatic light beams with opposite handedness of their circular polarization. The beams are in resonance with optical transitions in the Cs D$_1$ line ($\lambda$$\,\approx\,$$895$~nm). This configuration has already been known for observing CPT resonances of an increased contrast compared to a standard single-beam scheme. However, in contrast to previous studies, we use a scheme with two independent pump and probe beams, where the probe beam transmission is separately monitored. The experiments are carried out with a buffer-gas-filled $5$$\times$$5$$\times$$5$~mm$^3$ vapor cell. It is shown that the resonance's line shape acquires asymmetry which can be efficiently controlled by a microwave (Raman) phase between the beams. As a proof of concept, we study the way how this asymmetry can help to significantly mitigate the influence of ac Stark (light) shift on a long-term frequency stability of CPT clocks. The experimental verification is performed both with a distributed-Bragg-reflector (DBR) laser and a vertical-cavity surface-emitting laser (VCSEL). The latter has a particular importance for developing MACs. The results of experiments are in qualitative agreement with analytical theory based on a double $\Lambda$ scheme of atomic energy levels.
\end{abstract}

\maketitle

\section{\label{sec:1}Introduction}

Since its discovery in 1976 \cite{Alzetta1976,Arimondo1976}, coherent population trapping (CPT) has found numerous applications in laser spectroscopy \cite{Arimondo1996}, subrecoil laser cooling \cite{Aspect1988}, optical communications \cite{Hau1999,Yudin2013} and other fields. Nowadays, CPT is highly demanded in quantum metrology for developing miniature atomic clocks (MACs) and atomic magnetometers (see review \cite{Kitching2018}). CPT resonance is also known as a ``dark'' resonance since it leads to reduction in fluorescence of an atomic vapor cell.

CPT spectroscopy is an all-optical technique that allows to avoid application of a microwave cavity in contrast to another competing approach based on optical-microwave double resonance \cite{Batori2022}. This, in turn, provides a possibility for extreme miniaturization ($<\,$$100$~cm$^3$) and significant reduction in power consumption ($\ll\,$$1$~W) of a quantum device. There is a number of advanced CPT-based MACs which demonstrate frequency instability, characterized by the Allan deviation ($\sigma_y$), as low as $\sim\,$$10^{-11}$ at 1 s integration time and around 10$^{-12}$ at 24 hours \cite{Zhang2019,Skvortsov2020}. Further increase in atomic clocks' frequency stability, i.e. reduction in $\sigma_y$, would contribute to the progress in satellite and satellite-free navigation systems \cite{Fernandez2017,Hollberg2020}, including navigation in deep space \cite{Nydam2017}, remote sensing \cite{Chow2019}, and other existing and emerging applications. Since MACs satisfy small weight and power (SWaP) requirements, they are best suited for various space missions with the use of nanosatellites \cite{Chow2019,Warren2019}.

\begin{figure}[!t]
\includegraphics[width=1\linewidth]{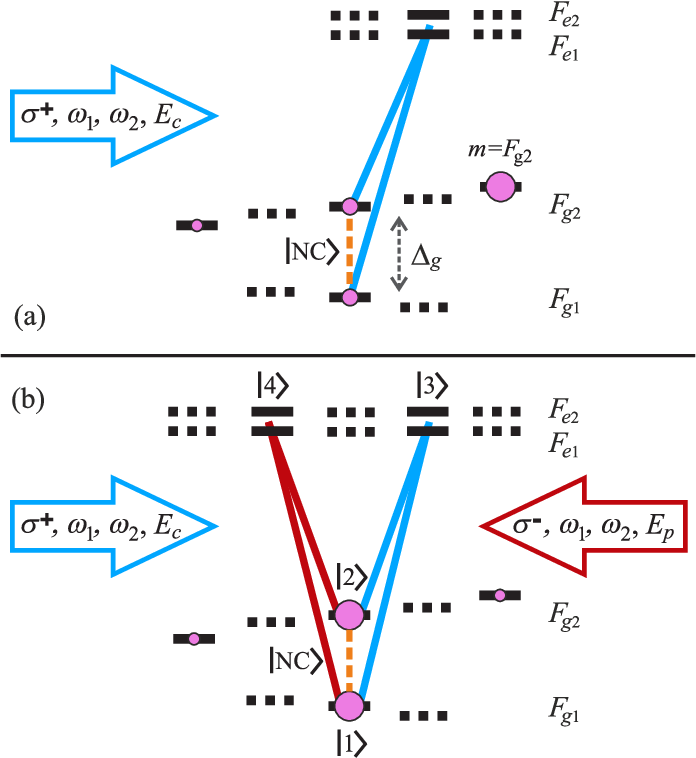}
\caption{\label{fig:1} Two light-field configurations for observation of CPT resonances in the D$_1$ line of an alkali-metal atom. Horizontal short solid and dashed lines denote Zeeman sub-levels. $\Delta_g$ is frequency of hyperfine splitting in the ground state. Pink circles qualitatively reflect the atomic populations of Zeeman (magnetic) sub-levels at $\delta_R$$\,=\,$$0$. The CPT (also known as noncoupled, ``NC'') state is shown as orange dashed line between two ground-state sub-levels with $m$$\,=\,$$0$. (a) A commonly used scheme, where blue oblique vertical lines denote $\sigma^+$ optical transitions ($\sigma\sigma$ scheme). (b) An excitation scheme with counter-propagating circularly polarized two-frequency beams, causing $\sigma^+$ (blue) and $\sigma^-$ (red) transitions ($\sigma^+\sigma^-$ scheme). Zeeman splitting under ambient longitudinal magnetic field (${\bf B}$$||$${\bf k}_{1,2}$) is shown only in the ground state. It is assumed that the hyperfine splitting in the excited state is not spectrally resolved due to collisional broadening of the optical transitions.}
\end{figure}

At short times ($\tau$$\,\sim\,$$1$$-$$100$~s), $\sigma_y$ is inversely proportional to a contrast-to-width ratio (CWR) \cite{Vanier2005}. In MACs, by using buffer gas in a microfabricated vapor cell, a full width at half maximum (FWHM) of CPT resonance typically takes a value of around 1~kHz. The resonance's contrast in a commonly used cw excitation scheme with a single multi-frequency circularly polarized beam (referred hear to as ``$\sigma\sigma$'' scheme) does not usually exceed $\approx\,$$1$~\% in microfabricated vapor cells (e.g., see \cite{Vicarini2018,Jia2022}). This drawback is explained by optical pumping of most atoms into a so-called ``stretch'' state, i.e. the Zeeman sub-level in atomic ground state with the highest or lowest magnetic quantum number $m$ [see Fig. \ref{fig:1}(a)]. Being in this state, the atoms do not interact with the resonant light at all, regardless of whether the Raman frequency detuning is equal to zero or not.


Many excitation methods have been proposed that demonstrate much better contrast of CPT resonance than in the standard $\sigma\sigma$ scheme. They can be divided into two groups: continuous wave (cw) and pulsed methods. Besides the relatively high contrast, most of the pulsed methods provide an improvement of long-term frequency stability of a CPT clock due to significant mitigation the influence of the ac Stark (light-shift) effect on frequency of a CPT resonance (e.g., see \cite{Zanon2005,Hafiz2018,Yudin2018,Shuker2019,Radnatarov2023}). However, from the point of view of miniaturization, cw methods are preferable against pulsed ones because the latter require the use of bulky acousto-optic or electro-optic modulators.

In cw excitation methods, the solution to the problem of low contrast usually comes down to choosing the right light-field geometry. In particular, the ``\textit{lin}$\perp$\textit{lin}'' \cite{Zanon2005,Yun2011,Yun2012}, ``\textit{lin}$||$\textit{lin}'' \cite{Taichenachev2005,Kazakov2005,Mikhailov2010,Boudot2009,Matsumoto2022} and ``$\sigma^+$$\sigma^-$'' \cite{Kargapoltsev2004,Taichenachev2004} polarization geometries have been proposed (see a brief review \cite{Zhong2014}). Push-pull optical pumping (PPOP) is another efficient technique that utilizes a Michelson interferometer to obtain alternating opposite circular polarizations \cite{Jau2004,Liu2013,Hafiz2015}. This scheme can be considered as a version of $\sigma^+$$\sigma^-$ configuration. Other PPOP schemes with polarization modulation by means of an electro-optic modulator (EOM) \cite{Yun2016} or a liquid-crystal polarization rotator \cite{Yun2017} seem bulky and consume additional power. Therefore, they cannot be directly integrated into MACs.

It should be noted that some of the above listed cw excitation schemes still cannot be used in MACs owing to their complexity or incompatibility with miniature vapor cells where a relatively high buffer gas pressure is required ($>\,$$100$~Torr). It concerns the \textit{lin}$\perp$\textit{lin} and \textit{lin}$||$\textit{lin} schemes, even if these schemes can demonstrate some unique features in low-pressure buffer-gas cells, such as complete immunity of the resonance frequency to fluctuations of ambient magnetic field  \cite{Matsumoto2022}.

The $\sigma^+$$\sigma^-$ scheme \cite{Kargapoltsev2004,Taichenachev2004} seems the most attractive to apply in MACs because it works well in high-pressure buffer-gas cells and requires minimum number of additional optical elements or even without such elements, if two vertical-cavity surface-emitting lasers (VCSELs) are used \cite{Shah2006}. However, despite high expectations from the $\sigma^+$$\sigma^-$ scheme, it has successfully implemented in cold-atom CPT clocks \cite{Liu2017,Elgin2019,Liu2022} rather than in MACs. This is not due to the problems with miniaturization. Indeed, several extremely miniature physics packages for atomic laser spectroscopy with two light beams have already been demonstrated (e.g., see \cite{Shah2006,Knappe2007}). The main reason consists in the fact that, in $\sigma^+$$\sigma^-$ scheme, a CPT resonance acquires large contrast at increased light intensity when the linewidth starts to suffer from the power broadening as well. It means that CWR does not surpass significantly that in the standard $\sigma\sigma$ scheme. The other mentioned cw methods demonstrate about the same limitations, leading to a short-term frequency stability close to that in the standard scheme \cite{Warren2017}. 


In this work, we show that the $\sigma^+$$\sigma^-$ configuration, besides an increased contrast of the resonance, can help to suppress the light-shift effect in CPT clocks. We consider a ``pump-probe'' configuration where the transmission of only the probe beam is monitored. The most previous studies were focused on analyzing the total absorption of both beams \cite{Kargapoltsev2004,Taichenachev2004,Shah2006,Affolderbach2002,Zhang2012}. The only exception is the work where two $\sigma^+$ and $\sigma^-$ beams were studied separately using a polarimeter after a $^{87}$Rb vapor cell \cite{Rosenbluh2006}. However, the authors proposed to use a differential channel of the polarimeter rather than a probe wave channel. Besides, they considered two light beams of equal intensities what did not allow to observe a sign-reversal effect, which we will demonstrate in our work. 

Here, we demonstrate that controllable asymmetry of the resonance's line shape in a probe beam transmission can be used for suppressing significantly the influence of optical as well as microwave power fluctuations on an error signal in CPT clocks. It should be emphasized that asymmetry of a CPT resonance can be observed even in the standard single-beam configuration \cite{Taichenachev2003}. However, this asymmetry has other physical reasons than that considered in our work and cannot be well controlled in MACs. In general, we believe that our proposal brings the $\sigma^+$$\sigma^-$ scheme back into the game, promising an improvement in long-term frequency stability of miniature CPT-based atomic clocks of a new generation.

\section{Theory}\label{SecTheory}

The $\sigma^+$$\sigma^-$ scheme can be arranged so that $\sigma^+$ and $\sigma^-$ beams are counter-propagating \cite{Affolderbach2002,Kargapoltsev2004,Taichenachev2004} or co-propagating \cite{Affolderbach2002,Shah2006,Zhang2012}. If the cell length is several times smaller than a half spatial period ($T_z$) of a microwave radiation at the frequency of atomic ground-state hyperfine splitting ($\Delta_g$), then these two schemes are equivalent. In particular, for Cs $\Delta_g$$\,\approx\,$$2\pi$$\times$$9.2$~GHz and we get $T_z$$/$$2$$\,=\,$$\pi$$c$$/$$\Delta_g$$\,\approx\,$$1.6$~cm. In our experiments, we use a 5 mm long cesium vapor cell and a scheme with counter-propagating beams.

In this section, we provide a brief theoretical treatment of light interaction with a closed double $\Lambda$-scheme, consisting of four Zeeman sub-levels $|1\rangle$$\dots$$|4\rangle$ as shown in Fig. \ref{fig:1}b. Each of the beams is composed of two monochromatic plane waves of the same strength, traveling along the quantization axis $z$:


\begin{eqnarray}\label{lightfield}
&&{\bf E}(z,t) = E_c\,{\bf e}_{+1} \Bigl[e^{-i(\omega_1t - k_1z)} + e^{-i(\omega_2t - k_2z)}\Bigr]\nonumber\\ &&+ E_p\,{\bf e}_{-1}\Bigl[e^{-i(\omega_1t + k_1z +\varphi_1)}\!+e^{-i(\omega_2t + k_2z +\varphi_2)}\Bigr]\!+c.c.,
\end{eqnarray}

\noindent where $E_c$, $E_p$ are the real amplitudes of control and probe beams, respectively, $k_{1,2}$ are the absolute values of wave vectors ($k_j$$\,=\,$$\omega_j/c$, with $c$ being the speed of light), ``\textit{c.c}'' means complex conjugate terms. In our experiment, phases $\varphi_{1,2}$ can be controlled by moving corresponding mirrors. Unit complex vectors of spherical basis ${\bf e}_{\pm1}$ describe opposite circular polarizations of the waves and cause optical $\sigma^+$ and $\sigma^-$ transitions in the atom.

We use the density matrix formalism according to quantum kinetic equation \cite{Rautian}:

\begin{equation}\label{DensityMatrix}
\frac{\partial}{\partial t}\,\hat{\rho} = -\frac{i}{\hbar}\Bigl[\bigl(\widehat{H}_0+\widehat{V}_e\bigr),\hat{\rho}\Bigr]-\widehat{\mathcal{R}}\bigl\{\hat{\rho}\bigr\}\,,
\end{equation}

\noindent where the square brackets $\bigl[\ldots,\ldots\bigr]$ stand for the commutation operation of two matrices, $\hat{H}_0$ is a part of the total Hamiltonian for a free atom, $\hat{V}_e$ describes the interaction between the atoms and the light field. The linear functional $\hat{\cal R}$ is responsible for the relaxation processes in the atom. Equation (\ref{DensityMatrix}) does not take into account motion of atoms in a gas that is a commonly used approximation in the case of buffer-gas-filled vapor cells.

We use the Dirac's bra $\langle \dots |$ and ket $|\dots\rangle$ vectors to present the density matrix in the form: 

\begin{equation}\label{BraKet}
   \hat{\rho} = \sum_{m = 1}^4 \sum_{n = 1}^4 \rho_{mn}\,|m\rangle \langle n|\,.
\end{equation}

\noindent Since the density matrix is Hermitian, $\hat{\rho}^\dag$$\,=\,$$\hat{\rho}$, we get the following relations between the matrix elements: $\rho_{nm}$$\,=\,$$\rho_{mn}^*$. In (\ref{BraKet}), the diagonal elements $\rho_{nn}$ are populations of magnetic sub-levels, the non-diagonal elements $\rho_{12}$ and $\rho_{21}$ are usually called microwave (or low-frequency) coherences, while $\rho_{13}$, $\rho_{14}$, $\rho_{23}$, $\rho_{24}$ and complex conjugate elements are known as optical coherences. We assume that the Zeeman coherences $\rho_{34}$ and $\rho_{43}$ are not created due to rapid collisional depolarization of the excited state. 

The free-atom Hamiltonian $\hat{H}_0$ in (\ref{DensityMatrix}) has diagonal form:

\begin{equation}\label{H0}
\hat{H}_0=\sum\limits_{n=1}^4{\cal E}_n|n\rangle \langle n|\,,
\end{equation}

\noindent where ${\cal E}_n$ is an energy of $n$ sub-level with ${\cal E}_3$$\,=\,$${\cal E}_4$, so that $\omega_{31}$$\,=\,$$\omega_{41}$$\,=\,$$\bigl({\cal E}_3$$-$${\cal E}_1\bigr)/\hbar$ and $\omega_{32}$$\,=\,$$\omega_{42}$$\,=\,$$\bigl({\cal E}_3$$-$${\cal E}_2\bigr)/\hbar$ being the optical transition frequencies in the double $\Lambda$ scheme.




The operator $\hat{V}_e$ in the rotating-wave and electro-dipole approximations reads


\begin{eqnarray}\label{Voperator}
\hat{V}_e=&&-\hat{{\bf d}}{\bf E}= -\hbar R_c e^{-i(\omega_1t-k_1z)} |3\rangle\langle 1|\nonumber\\
&& - \hbar R_c e^{-i(\omega_2t-k_2z)} |3\rangle\langle 2| \nonumber\\
&& -\hbar R_p e^{-i(\omega_1t+k_1z+\varphi_1)} |4\rangle\langle 1|\nonumber\\
&& -\hbar R_p e^{-i(\omega_2t+k_2z+\varphi_2} |4\rangle\langle 2| + H.c. \,.
\end{eqnarray}

\noindent Here, $\hat{{\bf d}}$ is the operator of atomic dipole moment, ``\textit{H.c.}'' means the Hermitian conjugate terms, and $R_{c,p}$$\,=\,$$d$$E_{c,p}$$/\hbar$ are the real positive values called the Rabi frequencies, with $d$ being the matrix element of dipole operator. This element is the same, in absolute value, for all four transitions in the considered double $\Lambda$-scheme. The latter is a feature of the real D$_1$ line. The sign of $d$, meanwhile, can be different for different transitions in the $\Lambda$-scheme owing to the properties of Clebsch-Gordan coefficients \cite{Varshalovich}. We assume that the signs of matrix elements are taken into account by the phases $\varphi_{1,2}$. 


The relaxation operator in (\ref{DensityMatrix}) can be divided into several parts:

\begin{equation}\label{RelParts}
   \widehat{\mathcal{R}} = \widehat{\mathcal{R}}_{therm} + \widehat{\mathcal{R}}_{sp} + \widehat{\mathcal{R}}_{col}\,,
\end{equation}

\noindent where relaxation of light-induced quantum state in the atom toward thermal equilibrium distribution of sub-level populations is described by the term:

\begin{equation}\label{GammaRel}
   \widehat{\mathcal{R}}_{therm} = \Gamma\hat{\rho} - \frac{\Gamma}{2}\sum_{n=1}^2 |n\rangle \langle n|\,,
\end{equation}

\noindent where we assume that initial (outside the light beam) ground-state sub-level populations are equal to $1/2$. $\Gamma$ determines a minimum linewidth of CPT resonance and is inversely proportional to the lifetime ($\tau$) of anisotropy in the atomic ground-state.

Let as present the spontaneous relaxation term in a matrix format:



\begin{equation}\label{SponRel}
   \widehat{\mathcal{R}}_{sp} = \gamma \left(
   \begin{array}{cccc}
       -\rho_{33}-\rho_{44} & 0 & \rho_{13} & \rho_{14} \\
       0 &  -\rho_{33}-\rho_{44} & \rho_{23} & \rho_{24} \\
       \rho_{31} & \rho_{32} & 2\rho_{33} & 0 \\
        \rho_{41} &  \rho_{42} &  0 & 2\rho_{44}
   \end{array}
   \right).
\end{equation}


\noindent Here, $\gamma$ is a half spontaneous relaxation rate ($2\gamma$$\,\approx\,$$4.57$~MHz for the Cs D$_1$ line \cite{Young1994}).

The collisional broadening of the optical transitions can be described by the following relaxation term:

\begin{equation}\label{CollRel}
   \widehat{\mathcal{R}}_{col} = \gamma_c \left(
   \begin{array}{cccc}
       0 & 0 & \rho_{13} & \rho_{14} \\
       0 &  0 & \rho_{23} & \rho_{24} \\
       \rho_{31} & \rho_{32} & 0 & 0 \\
        \rho_{41} &  \rho_{42} &  0 & 0
   \end{array}
   \right)\,,
\end{equation}

\noindent where $\gamma_c$ is the collisional relaxation rate which, in our case, is significantly larger than other relaxation rates mentioned above as well as the Doppler broadening of the optical transitions. The latter allows to do not take into account the motion of atoms.

As shown in Appendix, the probe wave absorption in a vapor cell is governed by the equation also known as the Beer-Lambert law:

\begin{equation}\label{BugerLaw}
    \frac{dI_p}{dz} = -\alpha\,I_p\,,
\end{equation}

\noindent with $I_p$ being the probe wave intensity and $\alpha$ being the absorption index. We calculated this coefficient analytically for the case when all light waves in (\ref{lightfield}) are in resonance with the corresponding optical transitions in the $\Lambda$ scheme (see Fig.\ref{fig:1}). We also assume that the following condition is satisfied:


\begin{equation}\label{xiparameter}
    \chi_{c,p} \ll 2\gamma\tau\,,
\end{equation}

\noindent where

\begin{equation}\label{chiparameter}
\chi_{c,p} = \frac{R^2_{c,p}}{\gamma_{eg}\Gamma}
\end{equation}

\noindent is the ratio between the optical pumping rate ($R^2_{c,p}$$/\gamma_{eg})$ and the ground-state relaxation rate ($\Gamma$).

After all the approximations have been made, we arrive at the following expression for the absorption index (see Appendix):

\begin{widetext}

\begin{equation}\label{Alpha}
    \alpha = \alpha_0 \Biggl[1-\frac{2\Delta\Gamma\chi_p}{\Delta^2+\delta^2_R}\Biggl\{1+\frac{\chi_c}{\chi_p}\cos{\psi}-\frac{\chi_c\delta_R}{\chi_p\Delta}\sin{\psi}\Biggr\}
    \Biggr]\,,
\end{equation}

\end{widetext}

\noindent where $\alpha_0$$\,=\,$$3\gamma\lambda^2 n_a/8\pi\gamma_{eg}$ is the index of linear absorption of the probe beam, i.e. the absorption index under the weak fields regime: $\chi_{c,p}$$\,\ll\,$$1$. Here $n_a$ is the atomic number density, $\gamma_{eg}$$\,=\,$$\gamma_c+\gamma+\Gamma$ is the relaxation rate of the optical coherences which determines a homogeneous broadening of the optical absorption line. The total phase $\psi$ in (\ref{Alpha}) reads: $\psi$$\,=\,$$2k_{12}z$$+$$\varphi_{12}$ with $k_{12}$$\,=\,$$k_1$$-$$k_2$, $\varphi_{12}$$\,=\,$$\varphi_1$$-$$\varphi_2$. We assume $z$ to be a parameter rather than a variable as in (\ref{BugerLaw}). Strictly speaking, this approach is valid only, if length of the cell ($L_{cell}$) is sufficiently small, so that $k_{12}L_{cell}$$\,\ll\,$$\pi$. This requirement can be easily satisfied in microfabricated (MEMS) vapor cells commonly used in MACs \cite{Vicarini2018}. To simplify the theoretical analysis, we also consider this assumption to be valid.

Eq. (\ref{Alpha}) contains a resonance feature, when the Raman (two-photon) frequency detuning, $\delta_R$$\,=\,$$\omega_1$$-$$\omega_2$$-$$\Delta_g$, is scanned around zero. Its half width at half maximum (HWHM) is $\Delta$$\,=\,$$\Gamma+2(R_c^2+R_p^2)/\gamma_{eg}$. This expression reflects a linear behavior of power broadening of the resonance that is typical for CPT resonances in buffer-gas cells \cite{Vanier2005}.

The derived absorption index reflects several interesting features of the dark resonance. In particular, the last term in curly brackets causes asymmetry of the resonance since it is an odd function in $\delta_R$. The same conclusion was made in \cite{Rosenbluh2006}. Another feature, missed in \cite{Rosenbluh2006}, concerns the effect of sign reverse of the resonance. Indeed, if the probe beam is weaker than the pump beam ($\chi_p$$\,<\,$$\chi_c$) and $\psi$$\,=\,$$\pi$, then the dark resonance is converted to the bright one that can be referred to as electromagnetically induced absorption (EIA). Such a resonance can be also considered for application in atomic clocks \cite{Brazhnikov2019}, however, this is outside the scope of the present work. 

\begin{figure}[!t]
\includegraphics[width=1\linewidth]{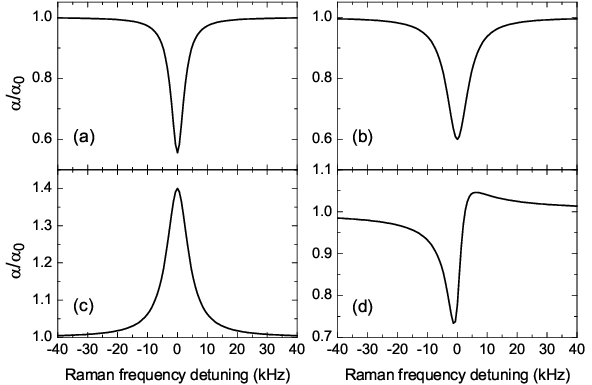}
\caption{\label{fig:2} Theoretical calculation of normalized absorption index of probe beam versus Raman frequency detuning: (a) $\psi$$\,=\,$$0$, $R_p$$\,=\,$$R_c$$\,=\,$$0.1\gamma$, (b) $\psi$$\,=\,$$\pi$, $R_p$$\,=\,$$2R_c$$\,=\,$$0.2\gamma$, (c) $\psi$$\,=\,$$\pi$, $R_p$$\,=\,$$0.5R_c$$\,=\,$$0.1\gamma$, (d) $\psi$$\,=\,$$\pi/2$, $R_p$$\,=\,$$R_c$$\,=\,$$0.1\gamma$.}
\end{figure}

Let us demonstrate graphically the main features of eq. (\ref{Alpha}). We first set typical values for the parameters: $\lambda$$\,=\,$$895$~nm, $2\gamma$$\,=\,$$2\pi$$\times$$4.57$~MHz, $\gamma_c$$\,=\,$$500$$\gamma$, $\Gamma$$\,=\,$$10^{-4}$$\gamma$, $n_a$$\,=\,$$2\times$$10^{11}$~cm$^{-3}$ (at $T_{cell}$$\,=\,$$40^\circ$C). Other parameters are written in caption of Fig. \ref{fig:2}. Under these conditions, the linear absorption index, $\alpha_0$, is around 0.38. Therefore, at $L_{cell}$$\,=\,$$5$~mm, the optical density (OD$\,=\,$$\alpha L_{cell}$) of the medium equals to $\approx\,$$0.19$, meaning that $\approx\,$$20\%$ of the light is absorbed in the cell out of the nonlinear resonance ($\delta_R$$\,\gg\,$$\Delta$).

If $\psi$$\,=\,$$0$, as in Fig. \ref{fig:2}a, the resonance is symmetric and represents a dark resonance regardless the ratio between strengths of the counter-propagating beams. The height of the resonance can be easily derived from (\ref{Alpha}):

\begin{equation}\label{Adark}
    A_{dark}=\alpha_0 \frac{2(\chi_c+\chi_p)}{1+2(\chi_c+\chi_p)}.
\end{equation}

\noindent When power broadening of the resonance prevails over relaxation rate of the ground state, i.e. $\chi_c$ or $\chi_p$ much larger than 1, the resonance's height tends to $\alpha_0$, meaning that the light absorption is very small at center of the resonance due to the CPT phenomenon. Note, that the derived expression for the resonance's height (\ref{Adark}) describes correctly the resonance's height in the transmitted light intensity only under the low optical density limit (OD$\,\ll$$1$). Otherwise, the differential equation (\ref{BugerLaw}) should be solved taking into account the dependence of $\alpha$ on the $z$ coordinate. For our quantitative treatment of the problem, it would be an unwanted complication.

Fig. \ref{fig:2}b shows the resonance curve at $\psi$$\,=\,$$\pi$ and $\chi_p$$\,>\,$$\chi_c$. The sign of the resonance is not changed. However, if $\chi_p$$\,<\,$$\chi_c$ (Fig. \ref{fig:2}c), a dark resonances is transformed to a bright resonance, which height follows from (\ref{Alpha}):

\begin{equation}
    A_{bright}=\alpha_0 \frac{2(\chi_c-\chi_p)}{1+2(\chi_c+\chi_p)}.
\end{equation}

\noindent It is now clear that, under the equal strengths of the beams ($\chi_p$$\,=\,$$\chi_c$) and $\psi$$\,=\,$$\pi$, a nonlinear resonance (either bright or dark) cannot be observed at all. This is the result of ``competition'' between the counter-propagating beams. Indeed, the pump beam ``tries'' to create some dark state in the atom that we denote as $|\text{NC}_1\rangle$. The probe beam does the same, but with the state $|\text{NC}_2\rangle$. The fact is that, at $\psi$$\,=\,$$\pi$, these two states are orthogonal, i.e. $\langle \text{NC}_1|\text{NC}_2 \rangle$$\,=\,$$0$. Such a situation can be also treated as a destructive interference of two two-photon transitions in the $\Lambda$-scheme and leads to the absence of any CPT state in the atom. The similar description in detail can be found in \cite{Kargapoltsev2004,Brazhnikov2019}, therefore, we do not pay much attention to this issue in the present work.

In the general case, when $\sin \psi$$\,\ne\,$$0$ in (\ref{Alpha}), the resonance acquires asymmetry (Fig. \ref{fig:2}d). In atomic clocks, the frequency detuning is modulated at some frequency to obtain the so-called ``error'' signal that is used to adjust a local oscillator frequency. Obviously, any asymmetry of the resonance's line shape gives rise to an effective light shift of the error signal. Therefore, researchers commonly consider such an asymmetry as an unwanted effect and develop methods to suppress its influence (e.g., see \cite{Yudin2023}). However, in the next section, we will demonstrate experimentally that, in our case, asymmetry can be controlled and can help to mitigate the light (ac Stark) shift effect on the error signal.

\section{Experiments}\label{sec:3}
\subsection{Scheme with a DBR laser}\label{subsecDBR}

\begin{figure}[b]
\includegraphics[width=1\linewidth]{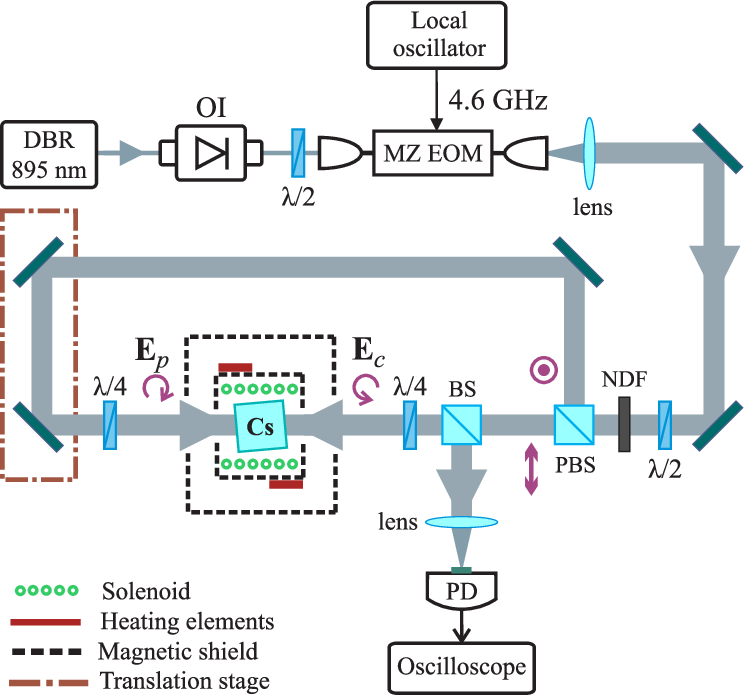}
\caption{\label{fig:3} Experimental setup for observation of CPT resonance in the Cs D$_1$ line in the $\sigma^+$$\sigma^-$ light-field configuration: DBR, distributed-Bragg-reflector laser diode; OI, optical isolator; $\lambda/2$, $\lambda/4$, half-wave and quarter-wave plates; MZ EOM, Mach-Zehnder electro-optic modulator; NDF, set of neutral density filters; PBS, polarizing beam splitter; BS, non-polarizing beam splitter; PD, photodetector.}
\end{figure}

Our theory considers a two-frequency light field, therefore, we first check the theoretical predictions with a single-mode DBR laser diode (Toptica Photonics, LD-0895-0040-DBR, $\Delta\nu$$\,\approx\,$$0.5$~MHz). A partial scheme of experimental setup is shown in Fig. \ref{fig:3}. A fiber-coupled Mach-Zehnder intensity electro-optic modulator (MZ EOM, iXblue Photonics, NIR-MX950-LN-20) allows to obtain only two optical frequency sidebands with angular frequencies $\omega_1$ and $\omega_2$. A home-made local oscillator drives the modulator at a microwave frequency of $\approx\,$$4.6$~GHz that is a half frequency of the hyperfine splitting in the Cs ground state ($\Delta_g$ in Fig. \ref{fig:1}a). The output collimator of MZ EOM and lens provide a laser beam of $\approx\,$$2$~mm in diameter ($1/e$).

The combination of a half-wave plate ($\lambda/2$), a gradient neutral density filter (NDF) and a polarizing beam splitter (PBS) provides smooth adjustment both the total light power in the vapor cell and the power of each of the counter-propagating beams. Quarter-wave plates ($\lambda/4$) are used to form the $\sigma^+$$\sigma^-$ light-field configuration in the cell. The cell has dimensions 5$\times$5$\times$5~mm$^3$ and is made of a borosilicate glass (``pyrex''). The cell is tilted to a small angle to prevent back scattering. It is filled with a buffer gas mixture Ar:Ne = 10:55 Torr. The optical frequency of laser radiation is stabilized with the help of an additional vapor cell with similar buffer gas pressure (not shown in Fig. \ref{fig:3}).

\begin{figure}[!t]
\includegraphics[width=1\linewidth]{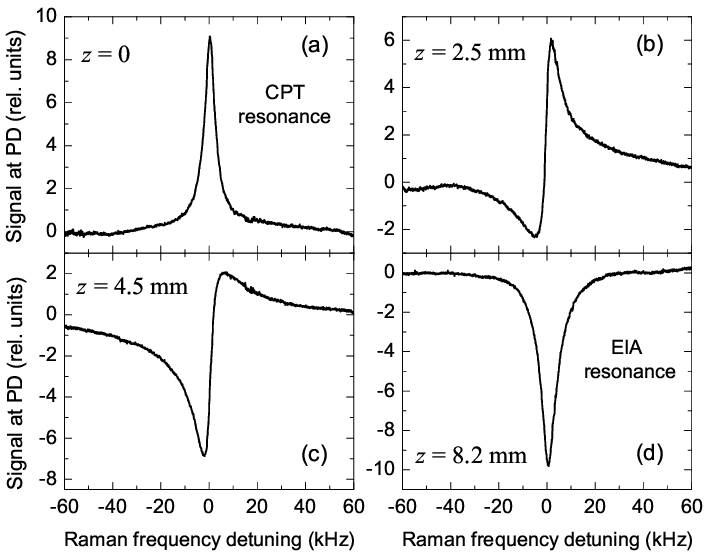}
\caption{\label{fig:4} Nonlinear resonance at different positions of translation stage with mirrors (see Fig. \ref{fig:3}). $P_c$$\,\approx\,$$390$~$\mu$W, $P_p$$\,\approx\,$$90$~$\mu$W. $T_{cell}$$\,\approx\,$$50^\circ$~C.}
\end{figure}

The vapor cell is placed inside a single-layer miniature magnetic shield with end caps made of $\mu$-metal. We use two transistors (Nexperia, PMZ290UNE2) soldered to the outer surface of the shield to heat the cell. This measure helps to mitigate the influence of a stray magnetic field from the heaters on CPT resonance \cite{Patent2019}. The NTC resistor is used as a thermosensor (not shown in the figure) to stabilize temperature of the package with $\sim\,$$1$~mK precision. The additional (external) 5-cm long magnetic shield is also used to suppress the laboratory magnetic field in the cell down to $\sim\,$1~mG. The miniature solenoid is installed inside the inner shield to produce a longitudinal magnetic field (${\bf B}$$||$${\bf k}$) in the cell of around 50~mG for shifting magnetic sub-levels in the ground state of cesium (see Fig. \ref{fig:1}). This allows to drive separately the so-called ``0-0'' two-photon transition in the atom, which serves as a ``clock'' transition. The non-polarizing beam splitter (BS) is used to direct the probe beam to a photodetector. Note that a PBS cannot be used for this purpose, since the probe beam has the same polarization as the pump beam at the considered point in the setup. 

\begin{figure}[!t]
\includegraphics[width=1\linewidth]{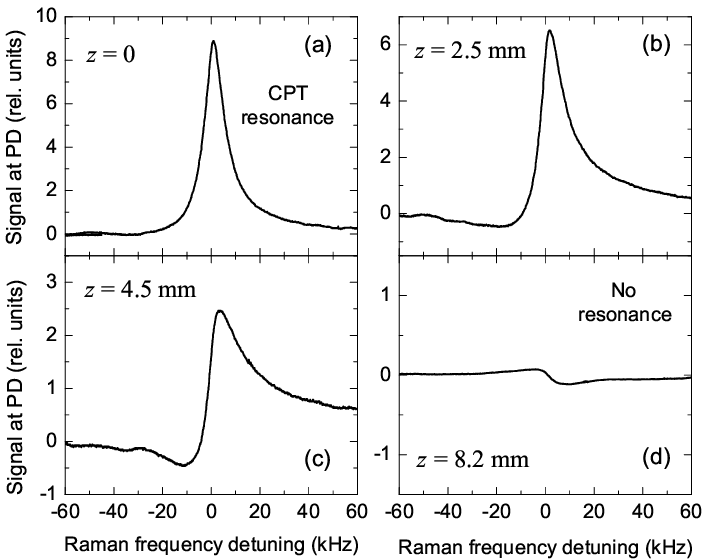}
\caption{\label{fig:5} Nonlinear resonance at different positions of translation stage with mirrors (see Fig. \ref{fig:3}). $P_c$$\,\approx\,$$P_p$$\,\approx\,$$370$~$\mu$W. $T_{cell}$$\,\approx\,$$50^\circ$~C.}
\end{figure}

A linear translation stage is used for accurate movement of two mirrors which direct the probe beam to the vapor cell. This stage allows changing the phases $\varphi_{1,2}$ in (\ref{lightfield}) and, ultimately, the total phase $\psi$ in (\ref{Alpha}). This parameter is used in our scheme to control a degree of asymmetry of CPT resonance. Indeed, as seen from Fig. \ref{fig:4}, if $I_p$$\,<\,$$I_c$, the nonlinear resonance is transformed from dark resonance (Fig. \ref{fig:4}a) to bright one (Fig. \ref{fig:4}d) just by changing position of the mirrors relative to the cell (we assume the $z$ coordinate of the stage to be equal to zero when a symmetric dark resonance is observed). However, if $I_p$$\,\approx\,$$I_c$, the resonance always stays of a dark type (Fig. \ref{fig:5}). These experimental results fit qualitatively the theoretical predictions (see Section \ref{SecTheory}). Particularly, there is almost no nonlinear resonance at some position of the mirrors (Fig. \ref{fig:5}d) as the result of destructive interference of the CPT states. A small nonlinear signal at center of the plot can be explained as the result of slight imbalance in the intensities of the beams.

\begin{figure}[!t]
\includegraphics[width=0.9\linewidth]{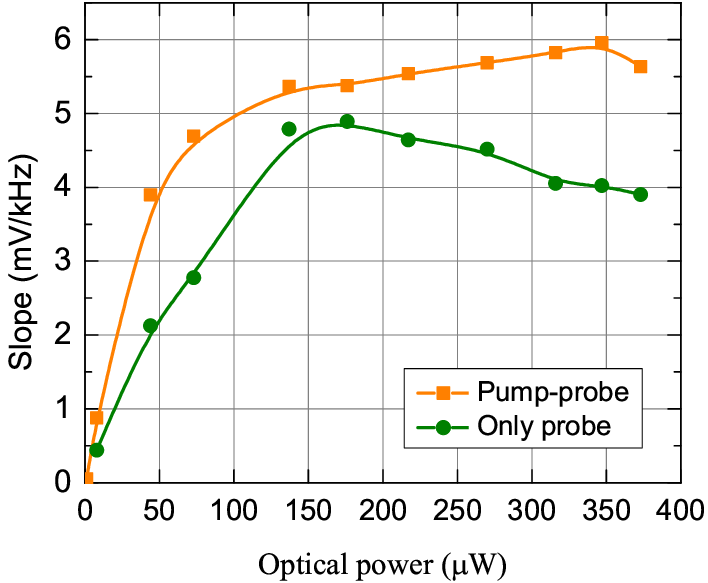}
\caption{\label{fig:6} Slope of error signal (a height-to-width ratio) in the $\sigma^+$$\sigma^-$ scheme (orange squares) and in the standard $\sigma\sigma$ scheme (green circles). $P_c$$\,\approx\,$$P_p$$\,\approx\,$$370$~$\mu$W, $T_{cell}$$\,\approx\,$$50^\circ$~C.}
\end{figure}

In the present work, we do not focus on studying the parameters of the observed resonances such as height, width and contrast. Instead, we show Fig. \ref{fig:6} to demonstrate some benefit from using the $\sigma^+$$\sigma^-$ scheme against the standard single-beam scheme in terms of the resonance slope. It is nothing but the height-to-width ratio of CPT resonance. This parameter determines a short-term frequency stability of an atomic clock \cite{Vanier2005}. As seen from the figure, the slope can be 20\% higher in the $\sigma^+$$\sigma^-$ scheme than in the standard scheme. This relatively small benefit does not make the $\sigma^+$$\sigma^-$ scheme much more attractive from point of view of short-term frequency stability of the clocks.

Let us now study a shift of error signal. To form this signal, we use a synchronous modulation/demodulation technique that is similar to the well-known Pound-Drever-Hall (PDH) technique for laser frequency stabilization by means of locking to a stable cavity (e.g., see \cite{Black2001}). One of the features of the classic PDH technique is that the frequency of modulation of the laser radiation is several times higher than the optical resonance linewidth. It has already been shown by several groups of researchers that the PDH-like technique has some advantages over the standard technique in the case of CPT clocks (\cite{BenAroya2007,Mikhailov2010,Yudin2017,Chuchelov2018,Yudin2023}).

\begin{figure}[!t]
\includegraphics[width=1\linewidth]{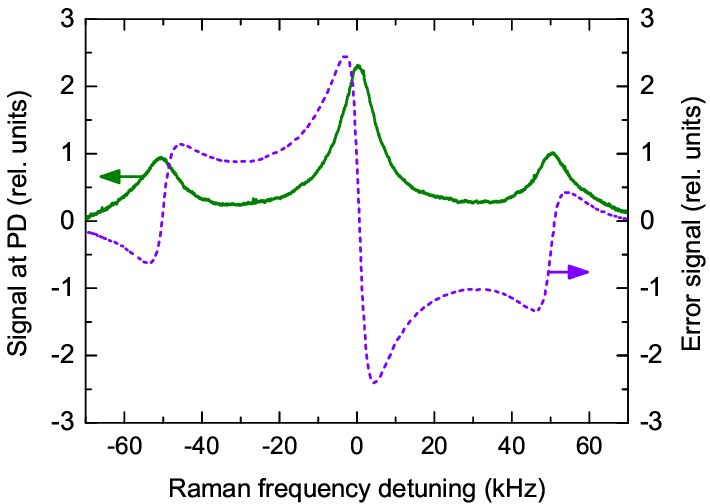}
\caption{\label{fig:7} CPT resonance in the case of $51$~kHz modulation of the Raman detuning (solid green line) and the corresponding error signal (dashed violet line). $P_c$$\,\approx\,$$P_p$$\,\approx\,$$300$~$\mu$W, $T_{cell}$$\,\approx\,$$50^\circ$~C.}
\end{figure}

In our experiments, LO frequency is modulated at 51~kHz, leading to observation of a three-peak CPT resonance on the oscilloscope (Fig. \ref{fig:7}, solid green curve). The corresponding error signal is also shown (dashed violet curve). Note that the buffer-gas collisional shift and quadratic Zeeman shift of the clock transition is subtracted from the Raman frequency detuning. The LO has a 10 MHz output that is used to measure the shift of the error signal with the help of a frequency comparator and a hydrogen microwave frequency standard (both provided by ``Vremya-Ch'' JSC \cite{VremyaCh}).

Fig. \ref{fig:8} shows the behavior of error signal shift versus the total light power in the cell controlled by NDF (see Fig. \ref{fig:3}). Note that in all experiments the cell temperature is relatively low, being in the range $\approx\,$$40$$-$$50^\circ$~C to make negligible the influence of optical density of the medium on the shift behavior \cite{Masian2015}. In the single-beam case the shift demonstrates the well-known linear law (black line with empty squares). However, in the $\sigma^+$$\sigma^-$ scheme the situation is changed qualitatively. Indeed, at some positions of the translation stage, the shift curve can change its sign or even acquire a smooth extremum. It means that, near this ``magic'' point, the error signal is almost immune to the small variations in optical power.




\subsection{Scheme with a VCSEL}\label{subsecVCSEL}

\begin{figure}[!t]
\includegraphics[width=1\linewidth]{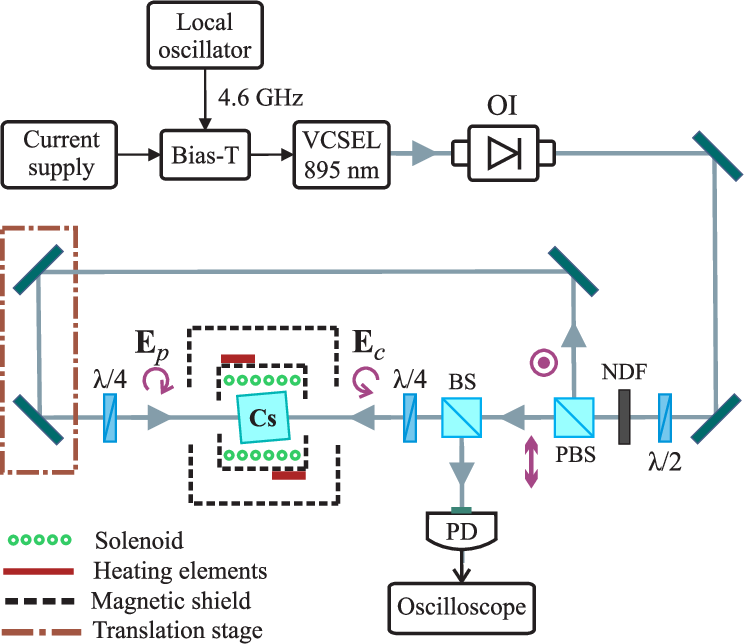}
\caption{\label{fig:9} Experimental setup for observation of CPT resonance in the Cs D$_1$ line in the $\sigma^+$$\sigma^-$ light-field configuration: VCSEL, vertical-cavity surface-emitting laser diode; other elements are the same as in Fig. \ref{fig:3}.}
\end{figure}

A VCSEL diode is commonly used in MACs because it supports direct modulation of electric current at microwave frequency to obtain required optical sidebands. The current modulation leads to frequency modulation (FM) of laser radiation, so that spectrum of radiation consists of several frequency sidebands rather than two sidebands as in the case of a pair of DBR laser and MZ EOM (Section \ref{subsecDBR}). In this way, here we study similar scheme (Fig. \ref{fig:9}) but with the use of a VCSEL diode ($\Delta\nu$$\,\approx\,$$50$~MHz) to verify the proposed light-shift compensation technique.

\begin{figure}[!b]
\includegraphics[width=1\linewidth]{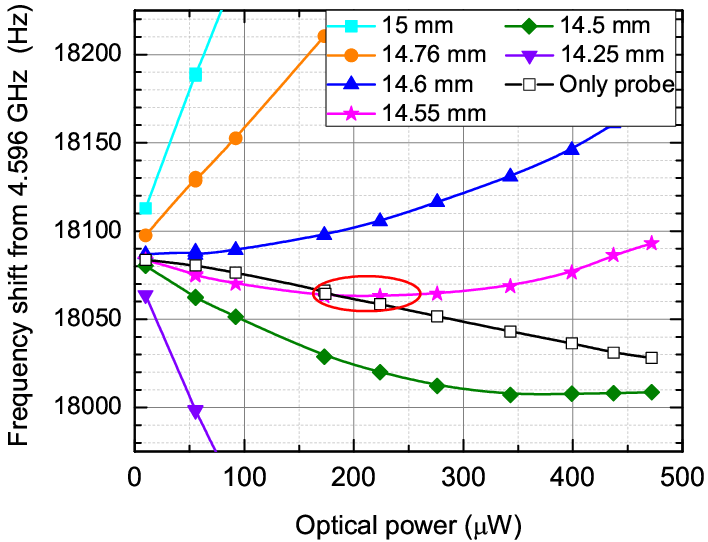}
\caption{\label{fig:8} Shift of error signal versus total optical power in the cell in the standard $\sigma\sigma$ scheme (black empty squares) and in the $\sigma^+$$\sigma^-$ scheme (other filled symbols). The optical powers of two beams are equal. $T_{cell}$$\,\approx\,$$40^\circ$~C.}
\end{figure}

\begin{figure}[!t]
\includegraphics[width=1\linewidth]{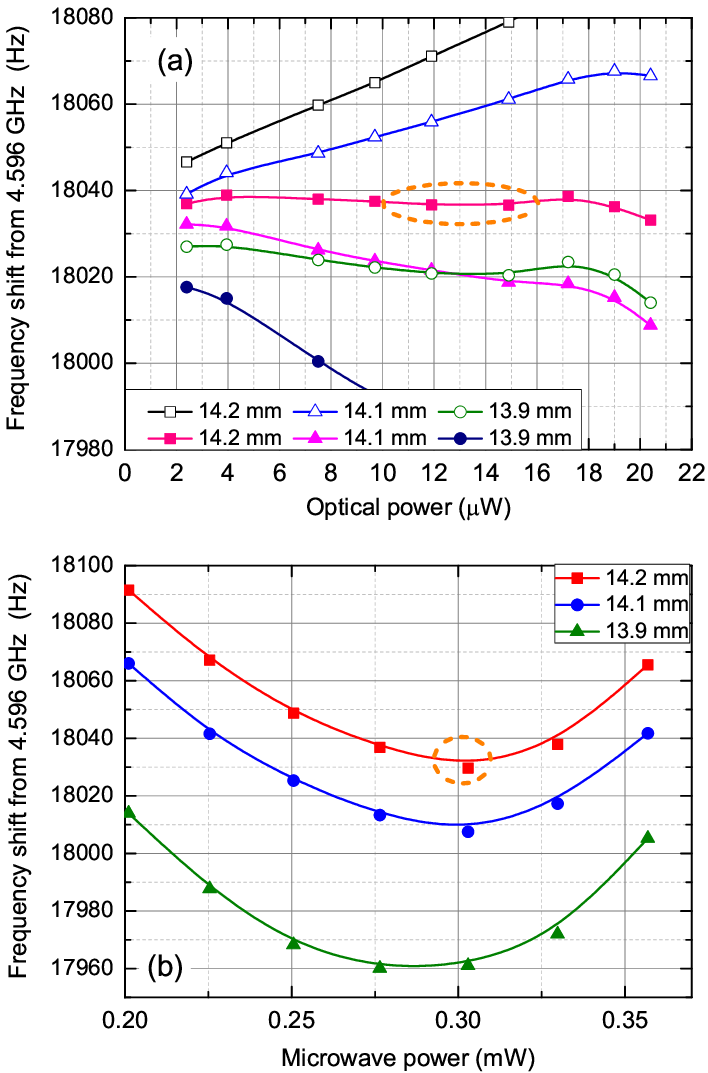}
\caption{\label{fig:10} Experiments with a VCSEL laser: shift of error signal at different positions of a translation stage (see Fig. \ref{fig:9}) versus total optical power (a) or LO microwave power (b). Filled symbols in (a) stand for optimal microwave power that can be figured out from (b) as an extremum of the corresponding curve (in vicinity of $0.3$~mW). Empty symbols in (a) correspond to maximum microwave power ($\approx\,$$0.37$~mW). Total optical power in (b) is equal to $\approx\,$$13$~$\mu$W. $T_{cell}$$\,\approx\,$$50^\circ$~C.}
\end{figure}

It is well-known that, in the case of VCSEL, the resonance shift behavior strongly depends on the microwave power $P_\mu$ supplied by LO. Namely, there is an optimal $P_\mu$ that provides a very low sensitivity of the error signal to optical power fluctuations (e.g., see \cite{Vanier2005,Mikhailov2010,Miletic2012}). In our case, the optimal microwave power depends on the phase $\psi$ as it seen from Fig. \ref{fig:10}a. In particular, at $z$$\,\approx\,$$14.2$~mm and $P_\mu$$\,\approx\,$$0.3$~mW (red curve with filled squares), the shift curve represents almost a horizontal line at a wide range of optical power. However, if $P_\mu$ is changed to $\,\approx\,$$0.37$~mW, the curve acquires a significant tilt (black curve with empty squares). A shift in vicinity of zero power in Fig. \ref{fig:10}a is equal to around $18030$~Hz. It differs from that in Fig. \ref{fig:8} ($\approx\,$$18080$~Hz), since we use different temperatures of cesium vapors in these two series of measurements as written in captions to the figures. In other words, it is a manifestation of a temperature shift of CPT resonance that, using the coefficients from \cite{Kozlova2011}, is estimated to be $\approx\,$$50$~Hz.

A brilliant feature of the considered excitation scheme is that the optimal microwave power, determined by Fig. \ref{fig:10}a, coincides with an extremum in Fig. \ref{fig:10}b at a curtain position of the translation stage (red squares). The curves in Fig. \ref{fig:10}a,b reveal existence of a special (``magic'') combination of values of the microwave modulation power, light field power and position of the translation stage (i.e. the phase $\psi$) that provides immunity of the error signal shift to fluctuations of both the optical and microwave power. Note that in other excitation schemes there has not been revealed such a unique combination of parameters. The only exception is the work \cite{Vaskovskaya2019} where, however, a buffer gas pressure should be controlled with a very high precision what is hardly possible during mass fabrication of MACs.




\section{Conclusions}\label{sec:4}

To conclude, we outline the ways for implementation of the proposed excitation scheme to MACs. The scheme provides a possibility for mitigation of influence of both the optical power and microwave power fluctuations on frequency stability of CPT-based atomic clocks. At the same time, before measuring the Allan deviation, the scheme requires a high degree of miniaturization of physics package as in chip-scale atomic clocks \cite{Kitching2018}. Indeed, there can be a non-negligible drift of the phase $\psi$ when the optical elements used in the setup are distributed over an optical table. This drift can be, for instance, caused by temperature variations and it can be a major limiting factor for achieving long-term frequency stability in a $10^{-13}$ range. In particular, in our bulky setup, position of the translation stage should be controlled with a $\sim\,$$100$~nm precision. In this sense, the results of our experiments can be considered as the proof of principle.

Despite the $\sigma^+$$\sigma^-$ scheme seems more complicated than the standard single-beam configuration commonly used in chip-scale CPT clocks, it can still be miniaturized to a large extent as it was shown, for instance, in the case of saturated-absorption rubidium spectrometer with counter-propagating pump-probe configuration \cite{Knappe2007}. The $\sigma^+$$\sigma^-$ scheme with co-propagating light beams can be more attractive \cite{Zhang2013}, because the phase $\psi$ in such a scheme can be controlled by only one optical element, a prism, which can be made of a material with low coefficient of thermal expansion.

Finally, an optimal way to solve the problem for controlling the Raman phase is to use two VCSELs mounted on the same substrate close to each other. Note that a mutual optical coherence of two laser fields is not required in our case. Therefore, there is no need for realization of an optical phase lock for these two lasers. Such a scheme with two VCSELs was proposed in \cite{Shah2006} for observation of high-contrast CPT resonances in the $\sigma^+$$\sigma^-$ configuration composed of co-propagating multi-frequency light beams. This scheme can be fabricated in a chip scale and provides a very fine tuning of the Raman phase $\psi$ just by means of electronics (a phase shifter) rather than with the help of mechanics. A key difference of our scheme is that a probe beam transmission should be monitored separately instead of monitoring a total light transmission through a vapor cell as in \cite{Shah2006}. This can be performed with the help of a polarimeter composed of a $\lambda/4$ plate and a polarizer. Alternatively, the polarimeter can be replaced by an integrated photonic spin selector \cite{Sebbag2021}.



The proposed scheme and the obtained results can be considered as the base for further detailed investigations in this direction that could lead to the development of a new-generation chip-scale atomic clock with a significantly improved long-term frequency stability below $10^{-12}$ at $24$~h.



\begin{acknowledgments}
The work was supported by Russian Science Foundation (Grant no. 22-12-00279).
\end{acknowledgments}

\appendix*

\section{Density matrix formalism}

Under the counter-propagating two-frequency light beams, the populations of magnetic sub-levels experience spatial oscillations owing to dependence of CPT state on the $z$ coordinate \cite{Brazhnikov2019}. In low-harmonic approximation, this leads to the following series expansion:

\begin{equation}\label{SeriesPopulations}
\rho_{nn}(z) = \rho_{nn}^{(0)} + \rho_{nn}^{(+)} e^{2i\,k_{12}z} +\rho_{nn}^{(-)}e^{-2i\,k_{12}z}\,,
\end{equation}

\noindent with $n$$\,=\,$$1,\dots,4$. The low-frequency coherences can be also expanded into the series:

\begin{eqnarray}
&&\rho_{12}(z,t) = e^{i\,\delta_{12}t}\Bigl[\rho_{12}^{(+)} e^{i\,k_{12}z} +\rho_{12}^{(-)}e^{-i\,k_{12}z}\Bigr]\,,\nonumber\\
&&\rho_{21}(z,t) = e^{-i\,\delta_{12}t}\Bigl[\rho_{21}^{(+)} e^{i\,k_{12}z} +\rho_{21}^{(-)}e^{-i\,k_{12}z}\Bigr]\,.\label{SeriesLowFreq}
\end{eqnarray}

\noindent Here $\delta_{12}$$\,=\,$$\omega_1$$-$$\omega_2$ and, since $\hat{\rho}$ is the Hermitian matrix, we get

\begin{equation}\label{ConjTerms1}
\rho_{21}^{(+)} = \rho_{12}^{(-)*}\,, \quad \rho_{21}^{(-)} = \rho_{12}^{(+)*}\,. 
\end{equation}

Analysis of density matrix equation (\ref{DensityMatrix}) with taking into account of (\ref{SeriesPopulations}) and (\ref{SeriesLowFreq}) leads to the following series expansion for the optical coherences:

\begin{eqnarray}\label{SeriesCoherences}
&&\rho_{13}(z,t) = e^{i\,\omega_1 t} \Bigl[\rho_{13}^{(-)} e^{-i\,k_1 z} +\rho_{13}^{(+12)}e^{i\,(k_1-2k_2)z}\Bigr]\,,\nonumber\\
&&\rho_{31}(z,t) = e^{-i\,\omega_1 t} \Bigl[\rho_{31}^{(+)} e^{i\,k_1 z} +\rho_{31}^{(-12)}e^{-i\,(k_1-2k_2)z}\Bigr]\,,\nonumber\\
&&\rho_{23}(z,t) = e^{i\,\omega_2 t} \Bigl[\rho_{23}^{(-)} e^{-i\,k_2 z} +\rho_{23}^{(-12)}e^{-i\,(2k_1-k_2)z}\Bigr]\,,\nonumber\\
&&\rho_{32}(z,t) = e^{-i\,\omega_2 t} \Bigl[\rho_{32}^{(+)} e^{i\,k_2 z} +\rho_{32}^{(+12)}e^{i\,(2k_1-k_2)z}\Bigr]\,,\nonumber\\
&&\rho_{14}(z,t) = e^{i\,\omega_1 t} \Bigl[\rho_{14}^{(+)} e^{i\,k_1 z+i\varphi_1} \nonumber\\
&&\qquad\qquad\qquad\qquad\qquad+\rho_{14}^{(-12)}e^{-i\,(k_1-2k_2)z+i\varphi_1}\Bigr]\,,\nonumber\\
&&\rho_{41}(z,t) = e^{-i\,\omega_1 t} \Bigl[\rho_{41}^{(-)} e^{-i\,k_1 z-i\varphi_1}\nonumber\\
&&\qquad\qquad\qquad\qquad\qquad+\rho_{41}^{(+12)}e^{i\,(k_1-2k_2)z-i\varphi_1}\Bigr]\,,\nonumber\\
&&\rho_{24}(z,t) = e^{i\,\omega_2 t} \Bigl[\rho_{24}^{(+)} e^{i\,k_2 z+i\varphi_2} \nonumber\\
&&\qquad\qquad\qquad\qquad\qquad+\rho_{24}^{(+12)}e^{i\,(2k_1-k_2)z+i\varphi_2}\Bigr]\,,\nonumber\\
&&\rho_{42}(z,t) = e^{-i\,\omega_2 t} \Bigl[\rho_{42}^{(-)} e^{-i\,k_2 z-i\varphi_2}\nonumber\\
&&\qquad\qquad\qquad\qquad+\rho_{42}^{(-12)}e^{-i\,(2k_1-k_2)z-i\varphi_2}\Bigr]\,,
\end{eqnarray}

\noindent where

\begin{equation}\label{ConjTerms2}
\rho_{nm}^{(+)} = \rho_{mn}^{(-)*}\,, \quad \rho_{nm}^{(+12)} = \rho_{mn}^{(-12)*}
\end{equation}

\noindent with $n,m$$\,=\,$$1\dots4$, $n$$\,\ne\,$$m$.

Based on the formulas (\ref{SeriesPopulations})-(\ref{SeriesCoherences}), we can derive from (\ref{DensityMatrix}) expressions for the spatial harmonics of optical coherences:

\begin{eqnarray}\label{OptCoherences}
&&\rho_{13}^{(-)} = i R_1 L_1^* \Bigl[\rho_{33}^{(0)} - \rho_{11}^{(0)}\Bigr] -i R_1 L_1^* \rho_{12}^{(-)}\,,\nonumber\\
&&\rho_{13}^{(+12)} = i R_1 L_1^* \Bigl[\rho_{33}^{(+)} - \rho_{11}^{(+)}\Bigr] -i R_1 L_1^* \rho_{12}^{(+)}\,,\nonumber\\
&&\rho_{23}^{(-)} = i R_1 L_2^* \Bigl[\rho_{33}^{(0)} - \rho_{22}^{(0)}\Bigr] -i R_1 L_2^* \rho_{21}^{(+)}\,,\nonumber\\
&&\rho_{23}^{(-12)} = i R_1 L_2^* \Bigl[\rho_{33}^{(-)} - \rho_{22}^{(-)}\Bigr] -i R_1 L_2^* \rho_{21}^{(-)}\,,\nonumber\\
&&\rho_{14}^{(+)} = i R_2 L_1^* \Bigl[\rho_{44}^{(0)} - \rho_{11}^{(0)}\Bigr] -i R_2 L_1^* \rho_{12}^{(+)}e^{-i\varphi_{12}}\,,\nonumber\\
&&\rho_{14}^{(-12)} = i R_2 L_1^* \Bigl[\rho_{44}^{(-)} - \rho_{11}^{(-)}\Bigr] -i R_2 L_1^* \rho_{12}^{(-)}e^{-i\varphi_{12}}\,,\nonumber\\
&&\rho_{24}^{(+)} = i R_2 L_2^* \Bigl[\rho_{44}^{(0)} - \rho_{22}^{(0)}\Bigr] -i R_2 L_2^* \rho_{21}^{(-)}e^{i\varphi_{12}}\,,\nonumber\\
&&\rho_{24}^{(+12)} = i R_2 L_2^* \Bigl[\rho_{44}^{(+)} - \rho_{22}^{(+)}\Bigr] -i R_2 L_2^* \rho_{21}^{(+)}e^{i\varphi_{12}}.
\end{eqnarray}

\noindent Other complex conjugate terms can be derived using relations (\ref{ConjTerms1}) and (\ref{ConjTerms2}). To obtain (\ref{OptCoherences}), we have introduced the complex Lorentzians

\begin{equation}\label{Lorentzians}
    L_{1,2} = \frac{1}{\gamma_{eg}-i\delta_{1,2}}
\end{equation}

\noindent with optical frequency detunings $\delta_1$$\,=\,$$\omega_1$$-$$\omega_{31}$$\,=\,$$\delta_0$$+$$\delta_R/2$, $\delta_2$$\,=\,$$\omega_2$$-$$\omega_{32}$$\,=\,$$\delta_0$$-$$\delta_R/2$. Here $\delta_0$ is the optical frequency detuning of the mean laser frequency from the mean transition frequency, i.e. $\delta_0$$\,=\,$$(\omega_1$$+$$\omega_{2})/2$$-$$(\omega_{31}$$+$$\omega_{32})/2$. In particular, if $\omega_1$ and $\omega_2$ are obtained in experiments by means of EOM, then they are optical sidebands of the $\pm1$ orders, while $\omega_0$$\,=\,$$(\omega_1$$+$$\omega_2)/2$ is the carrier frequency. To simplify our theory, we consider the case when $\delta_0$$\,=\,$$0$. Therefore, as seen from (\ref{Lorentzians}), $L_1$$\,=\,$$L_2^*$$\,\equiv\,$$L$.


We consider a closed system of energy levels (Fig. \ref{fig:1}). It means that the total sub-level population is constant and equals to unity:

\begin{equation}
    \rho_{11}+\rho_{22}+\rho_{33}+\rho_{44}=1\,.
\end{equation}

\noindent This equation results in three equations for spatial harmonics of sub-level populations:

\begin{eqnarray}
&&\rho_{11}^{(0)}+\rho_{22}^{(0)}+\rho_{33}^{(0)}+\rho_{44}^{(0)}=1\,,\nonumber\\
&&\rho_{11}^{(+)}+\rho_{22}^{(+)}+\rho_{33}^{(+)}+\rho_{44}^{(+)}=0\,,\nonumber\\
&&\rho_{11}^{(-)}+\rho_{22}^{(-)}+\rho_{33}^{(-)}+\rho_{44}^{(-)}=0\,.\label{HarmNorm}
\end{eqnarray}

Frequent collisions between the work alkali atoms and the buffer-gas atoms lead to total depolarization of the excited state, i.e. we can consider $\rho_{44}$$\,=\,$$\rho_{33}$. Besides, for the considered scheme of levels (Fig. \ref{fig:1}) and the Rabi frequencies, we have a symmetry of the density matrix equation with respect to the substitution $\rho_{11}$$\,\longleftrightarrow\,$$\rho_{22}$. This means nothing by the equality of these elements, i.e. $\rho_{11}$$\,=\,$$\rho_{22}$. Then, from (\ref{HarmNorm}) we get

\begin{eqnarray}\label{MainEqs0}
&&\rho_{22}^{(0)}=\rho_{11}^{(0)}\,,\quad\rho_{22}^{(+)}=\rho_{11}^{(+)}\,,\quad\rho_{22}^{(-)}=\rho_{11}^{(-)}\,,\nonumber\\
&&\rho_{33}^{(0)}=\frac{1}{2} - \rho_{11}^{(0)}\,,\quad
\rho_{33}^{(+)}=-\rho_{11}^{(+)}\,,\quad
\rho_{33}^{(-)}=-\rho_{11}^{(-)}\,,\nonumber\\
&&\rho_{44}^{(0)}=\rho_{33}^{(0)}\,,\quad
\rho_{44}^{(+)}=\rho_{33}^{(+)}\,,\quad
\rho_{44}^{(-)}=\rho_{33}^{(-)}\,.
\end{eqnarray}

\noindent Therefore, only harmonics of $\rho_{11}$ and $\rho_{12}$ should be derived to find all elements of the density matrix.

After taking into account all above, we obtain from (\ref{DensityMatrix}) the following set of equations:

\begin{eqnarray}\label{MainEqs1}
&&\Bigl( \Gamma + 2\gamma + 4\gamma_{eg}\bigl[S_c+S_p\bigr] \Bigr) \rho_{11}^{(0)} = \frac{1}{2}\bigl(\Gamma+2\gamma\bigr)\nonumber\\
&&\quad\quad+\gamma_{eg}\bigl(S_c+S_p\bigr) - R_c^2L\rho_{21}^{(+)} - R_c^2L^*\rho_{12}^{(-)}\nonumber\\
&&\quad\quad\quad-R_p^2L\rho_{21}^{(-)}e^{i\varphi_{12}} - R_p^2L^*\rho_{12}^{(+)}e^{-i\varphi_{12}}\,,\nonumber\\
&&\Bigl( \Gamma + 2\gamma + 2R_c^2L^* + 2R_p^2L \Bigr) \rho_{11}^{(+)} =\nonumber\\
&&\quad\quad\quad\quad\quad\quad-R_c^2L^*\rho_{12}^{(+)}-R_p^2L\rho_{21}^{(+)}e^{i\varphi_{12}}\,,\nonumber\\
&&\Bigl( \Gamma + 2\bigl[ R_c^2 + R_p^2 \bigr] + i\delta_R \Bigr) \rho_{12}^{(+)} = R_p^2L^*e^{i\varphi_{12}}\nonumber\\
&&\quad\quad\quad\quad\quad -4R_c^2L^*\rho_{11}^{(+)}-4R_p^2L^*\rho_{11}^{(0)}e^{i\varphi_{12}}\,,\nonumber\\
&&\Bigl( \Gamma + 2\bigl[ R_c^2 + R_p^2 \bigr] + i\delta_R \Bigr) \rho_{12}^{(-)} = R_c^2L^*\nonumber\\
&&\quad\quad\quad\quad\quad -4R_p^2L^*\rho_{11}^{(-)}e^{i\varphi_{12}}-4R_c^2L^*\rho_{11}^{(0)}\,.
\end{eqnarray}

\noindent The equations for $\rho_{11}^{(-)}$, $\rho_{21}^{(+)}$ and $\rho_{21}^{(-)}$ can be derived by complex conjugation of the equations for $\rho_{11}^{(+)}$, $\rho_{12}^{(-)}$ and $\rho_{12}^{(+)}$, respectively. Here the following saturation parameters have been used:

\begin{equation}
    S_i = \frac{R_i^2}{\gamma_{eg}^2+\bigl(\delta_R/2\bigr)^2}\,,
\end{equation}

\noindent with $i$$\,=\,$$c,p$.

In buffer-gas cells, optical absorption lines are considerably broadened and the CPT resonance is observed at a very small Raman detuning compared to $\gamma_{eg}$, i.e. $\delta_R$$\,\ll\,$$\gamma_{eg}$. Therefore, we can neglect $\delta_R$ in $L$ and $S$, considering $L$$\,\approx\,$$\gamma_{eg}^{-1}$, $S_i$$\,\approx\,$$R_i^2$$/\gamma_{eg}^2$ ($i$$\,=\,$$1,2$) in (\ref{MainEqs1}). Similarly to ``optical'' Lorentzians (\ref{Lorentzians}), we can introduce the two-photon Lorentzian:

\begin{equation}\label{RamanLorentzian}
    L_R = \frac{1}{\widetilde{\Delta} - i\widetilde{\delta}_R}\,,
\end{equation}

\noindent where resonance's HWHM and Raman detuning are expressed in $\Gamma$ units: $\widetilde{\Delta}$$\,=\,$$\Delta/\Gamma$, $\widetilde{\delta}_R$$\,=\,$$\delta_R/\Gamma$. Then, using the notation ({\ref{chiparameter}), the set of equations (\ref{MainEqs1}) can be rewritten in the following form:

\begin{eqnarray}
&&\bigl(\widetilde{\Delta} + 2(\chi_c+\chi_p) + 2\gamma\tau\bigr)\rho_{11}^{(0)} = \frac{1}{2}\bigl(\widetilde{\Delta}+2\gamma\tau\bigr)\nonumber\\
&&\quad\quad-\chi_c\text{Re}\Bigl\{\rho_{12}^{(-)}\Bigr\}-\chi_p\text{Re}\Bigl\{\rho_{12}^{(+)}e^{-i\varphi_{12}}\Bigr\}\,,\label{MainEqs21}\\ 
&&\bigl(\widetilde{\Delta} + 2\gamma\tau\bigr)\rho_{11}^{(-)} = -\chi_c\rho_{21}^{(-)}-\chi_p\rho_{12}^{(-)}e^{-i\varphi_{12}}\,,\label{MainEqs22}\\ 
&&\rho_{12}^{(-)} = \chi_c L_R^* \nonumber\\
&&\quad\quad\quad\quad - 4\chi_c L_R^*\rho_{11}^{(0)} - 4\chi_p L_R^*\rho_{11}^{(-)}e^{i\varphi_{12}}\,,\label{MainEqs23}\\ 
&&\rho_{12}^{(+)} = \chi_p L_R^*e^{i\varphi_{12}}\nonumber\\
&&\quad\quad\quad\quad - 4\chi_p L_R^*\rho_{11}^{(0)}e^{i\varphi_{12}} - 4\chi_c L_R^*\rho_{11}^{(+)}\,.\label{MainEqs24}
\end{eqnarray}

\noindent Here $\text{Re}\bigl\{\dots\bigr\}$ stands for a real part of complex number.

In alkali-atom vapor cells with buffer gas pressure of around 100 Torr and higher, the ground-state relaxation rate ($\Gamma$) is of the order of $10^{-5}$$-$$10^{-4}$$\gamma$ (depending on other parameters such, for instance, as diameter of laser beam). It means that, in (\ref{MainEqs21}) and (\ref{MainEqs22}), $\gamma\tau$$\,\sim\,$$10^4$$-$$10^5$. Meanwhile, in atomic clocks, a power broadening of the CPT resonance is not very high, i.e. $\chi_{c,p}$ and $\widetilde{\Delta}$ commonly lie in the range $1$$-$$10$. Therefore, condition (\ref{xiparameter}) is satisfied with a good margin, meaning that the terms proportional to $\gamma\tau$ have major impacts to the equations (\ref{MainEqs21}) and (\ref{MainEqs22}). In this work, it is enough to consider only these terms to explain qualitatively main features of CPT resonances observed in the experiments. Alternatively, one may use the perturbation theory approach to obtain more accurate expressions in the form of a power series expansion in the strength $\xi$$\,=\,$$\chi_{c,p}$$/$$2\gamma\tau$ of the perturbation.


Under the approximation made, almost all atomic populations are concentrated in the ground-state sub-levels $\rho_{11}$ and $\rho_{22}$, while spatial harmonics of these density matrix elements can be neglected. We can finally write the following approximate solution of the system (\ref{MainEqs0}), (\ref{MainEqs21})-(\ref{MainEqs24}):

\begin{eqnarray}
    \rho_{11}^{(0)} = \rho_{22}^{(0)} \approx \frac{1}{2}\,,\label{solution1}\\
    \rho_{11}^{(-)} = \rho_{11}^{(+)} = \rho_{22}^{(-)} = \rho_{22}^{(+)} \approx 0\,,\label{solution2}\\
     \rho_{33}^{(0)} = \rho_{44}^{(0)} \approx 0\,,\label{solution3}\\
      \rho_{33}^{(-)} = \rho_{33}^{(+)} = \rho_{44}^{(-)} = \rho_{44}^{(+)} \approx 0\,,\label{solution4}\\
    \rho_{12}^{(-)} \approx -\chi_c L_R^* \,,\label{solution5}\\
    \rho_{12}^{(+)} \approx -\chi_p L_R^*e^{i\varphi_{12}} \,,\label{solution6}\\
    \rho_{21}^{(-)} = \rho_{12}^{(+)*} \approx -\chi_p L_Re^{-i\varphi_{12}} \,,\label{solution7}\\
    \rho_{21}^{(+)} = \rho_{12}^{(-)*} \approx -\chi_c L_R \,.\label{solution8}
\end{eqnarray}

\noindent It is seen that after neglecting influence of spatial harmonics, such as $\rho_{11}^{(-)}$ and $\rho_{11}^{(+)}$, the CPT state in the atom can be created by each of the counter-propagating light beams independently, because any interference terms proportional to $\chi_c\chi_p$ vanish in (\ref{solution5})-(\ref{solution8}). Also, it can be easily shown that if the two-photon resonance condition is satisfied ($\delta_R$$\,=\,$$0$), one of the beams is absent (e.g., $\chi_p$$\,=\,$$0$) and the ground-state relaxation is absent as well, then the set of equations (\ref{solution1})-(\ref{solution8}) gives us the following simple solution: $\rho_{11}^{(0)}$$\,=\,$$\rho_{22}^{(0)}$$\,=\,$$1/2$, $\rho_{12}^{(-)}$$\,=\,$$\rho_{21}^{(+)}$$\,=\,$$-1/2$. This result is well known from ``classic'' theory of the CPT phenomenon based on a single $\Lambda$-scheme of atomic levels \cite{Arimondo1996}.

Evolution of amplitude of an electromagnetic wave in the vapor cell due to absorption obeys the equation:

\begin{eqnarray}\label{WaveEq1}
&&\frac{dE_{p1}(z)}{dz} = i 2\pi k_1 n_a d \rho_{41}(z,t)e^{i(\omega_1t+k_1z+\varphi_1)} =\nonumber\\
&&\quad\quad\quad i 2\pi k_1 n_a d \Bigl[\rho_{41}^{(-)} + \rho_{41}^{(+12)}e^{2ik_{12}z}\Bigr]\,,
\end{eqnarray}

\noindent where $E_{p1}$ is a slowly varying amplitude of one of the probe waves with frequency $\omega_1$ in (\ref{lightfield}), which drives the transition $|1\rangle$$\to$$|4\rangle$ (see Fig. \ref{fig:1}). To derive (\ref{WaveEq1}), we have used the expansion for $\rho_{41}(z,t)$ from (\ref{SeriesCoherences}). Taking into account (\ref{OptCoherences}) and (\ref{solution1})-(\ref{solution8}), we obtain

\begin{equation}\label{WaveEq2}
\frac{dE_{p1}}{dz} = -\frac{\pi k_1 n_a d R_p}{\gamma_{eg}}\Bigl[1-2L_R\bigl(\chi_p+\chi_ce^{i\psi}\bigr)\Bigr]\,.
\end{equation}

The other probe wave with amplitude $E_{p2}$ and frequency $\omega_2$ obeys the same equation as $E_{p1}$. The only difference is that we should take $k_2$ instead of $k_1$ in coefficient before the square brackets in (\ref{WaveEq2}). However, we can neglect negligible difference between the values $k_1$ and $k_2$ in the coefficient, so that we will further assume $k_1$$\,=\,$$k_2$$\,\equiv\,$$k$. Note that we still keep $k_{12}$$\,=\,$$k_1$$-$$k_2$$\,\ne\,$$0$ in $e^{i\psi}$ where $\psi$$\,=\,$$2k_{12}z$$+$$\varphi_{12}$.

The total intensity ($I_p$) of the probe light beam is determined by the expression:

\begin{equation}\label{IntensityDef}
    I_p = \frac{c}{2\pi} \overline{\Bigl|{\bf E}_{p1}(z,t)+{\bf E}_{p2}(z,t)\Bigr|^2}\,,
\end{equation}

\noindent where $\bigl|\dots\Bigr|$ stands for absolute value of a complex number, horizontal line over the formula means time averaging, and ${\bf E}_{p1}(z,t)$$\,=\,$$E_{p1}$${\bf e}_{-1}$$e^{-i(\omega_1t+k_1z+\varphi_1)}$, ${\bf E}_{p2}(z,t)$$\,=\,$$E_{p2}$${\bf e}_{-1}$$e^{-i(\omega_2t+k_2z+\varphi_2)}$. Let us remind that in our theory we consider $E_{p1}$$\,=\,$$E_{p2}$$\,\equiv\,$$E_p$ according to (\ref{lightfield}). Besides, we can neglect interference term $\sim$${\bf E}_{p1}$${\bf E}_{p2}^*$ in (\ref{IntensityDef}), because it oscillates in time at hyperfine frequency $\omega_{12}$ and is averaged in the experiments. Therefore, the probe beam intensity is $I_p$$\,=\,$$(c/\pi)E_{p1}^2$. Based on (\ref{WaveEq2}), we get

\begin{eqnarray}\label{WaveEq3}
    \frac{dI_p}{dz} &&= -\frac{2 k c n_a d E_p R_p}{\gamma_{eg}}\times\nonumber\\
    &&\Bigl(1-2\chi_p\text{Re}\bigl\{L_R\bigr\}-2\chi_c\text{Re}\bigl\{L_Re^{i\psi}\bigr\}\Bigr)\,.
\end{eqnarray}

\noindent Since $d^2$$\,=\,$$3\gamma\hbar/4k^3$, $R_p$$\,=\,$$dE_p/\hbar$ and $k$$\,=\,$$2\pi/\lambda$, taking into account (\ref{RamanLorentzian}), eq. (\ref{WaveEq3}) leads to

\begin{widetext}

\begin{equation}
\frac{dI_p}{dz} = -\frac{3\gamma n_a \lambda^2}{8\pi\gamma_{eg}}
\Biggl(1-\frac{2\chi_p\Gamma\Delta}{\Delta^2+\delta_R^2}\Bigl[1+\frac{\chi_c}{\chi_p}\cos{\psi} - \frac{\chi_c\delta_R}{\chi_p\Delta}\sin{\psi}\Bigr]\Biggr)\times I_p\,.
\end{equation}

\end{widetext}

\noindent The latter expression corresponds to (\ref{BugerLaw}) with the nonlinear absorption index (\ref{Alpha}).

\end{document}